\documentclass{article}

\usepackage[final]{ml4eng_2020}

\usepackage[colorlinks=true,bookmarksopen,pdfsubject={algorithms},linkcolor={blue}, anchorcolor={black}, citecolor={blue}, filecolor={magenta}, menucolor={black}, pagecolor={red},backref=none,urlcolor={blue}]{hyperref}

\usepackage[utf8]{inputenc} % allow utf-8 input
\usepackage[T1]{fontenc}    % use 8-bit T1 fonts
\usepackage{url}            % simple URL typesetting
\usepackage{booktabs}       % professional-quality tables
\usepackage{amsfonts}       % blackboard math symbols
\usepackage{nicefrac}       % compact symbols for 1/2, etc.
\usepackage{microtype}      % microtypography
\usepackage{bm}
\usepackage{enumitem}
\usepackage{graphicx}  
\usepackage{amsmath}
\usepackage{subfigure}
\usepackage[numbers, sort]{natbib}

\usepackage{lineno}
\usepackage{comment}
\usepackage{amsmath}
\usepackage{tabu}
\usepackage{setspace}
\usepackage{bm}
\usepackage{appendix}
\usepackage[utf8]{inputenc} % allow utf-8 input
\usepackage[T1]{fontenc}    % use 8-bit T1 fonts
\usepackage{url}            % simple URL typesetting
\usepackage{booktabs}       % professional-quality tables
\usepackage{amsfonts}       % blackboard math symbols
\usepackage{nicefrac}       % compact symbols for 1/2, etc.
\usepackage{microtype}      % microtypography
\usepackage{microtype}
\usepackage{mathrsfs}
\usepackage{graphicx}
\usepackage{subfigure}
\usepackage{amssymb}
\usepackage{booktabs} % for professional tables
\usepackage{bm}

\usepackage{bm}
\usepackage{amsmath}
\usepackage{amssymb}
\usepackage{setspace}
\usepackage{bigints}
\usepackage{array}
\usepackage{multirow}
\usepackage{colortbl}
\usepackage{graphicx}
\usepackage{subfigure}
\usepackage[table,xcdraw]{xcolor}
\usepackage{booktabs}
\usepackage{tabu}
\usepackage{bbm}
\usepackage{float}
\usepackage[singlelinecheck=false]{caption}
\usepackage[utf8]{inputenc}
\usepackage[english]{babel}
\usepackage{bookmark}
\usepackage{mathrsfs}
\usepackage{color}
\usepackage{mathtools} 
\usepackage{amssymb}
\usepackage{xfrac}
\usepackage{lscape}
\usepackage{pdflscape}
\usepackage{algorithm}% http://ctan.org/pkg/algorithm
\usepackage{algpseudocode}% http://ctan.org/pkg/algorithmicx
\usepackage{xspace}
\usepackage{booktabs}

\title{Thermodynamic Consistent Neural Networks for Learning Material Interfacial Mechanics}

\author{Jiaxin Zhang \\
  Computer Science and Mathematics Division\\
  Oak Ridge National Laboratory, Oak Ridge, TN 37830 \\
  \texttt{zhangj@ornl.gov} \\
   \And
   Congjie Wei \\
   Department of Civil, Architectural, and Environmental Engineering  \\
   Missouri University of Science and Technology, Rolla, MO 65409 \\
  \texttt{cw6ck@mst.edu} \\
   \AND
   Chenglin Wu \\
   Department of Civil, Architectural, and Environmental Engineering  \\
   Missouri University of Science and Technology, Rolla, MO 65409 \\
  \texttt{wuch@mst.edu} \\
}

\begin{document}

\maketitle

\begin{abstract}
For multilayer materials in thin substrate systems, interfacial failure is one of the most challenges. The traction-separation relations (TSR) quantitatively describe the mechanical behavior of a material interface undergoing openings, which is critical to understand and predict interfacial failures under complex loadings. However, existing theoretical models have limitations on enough complexity and flexibility to well learn the real-world TSR from experimental observations. A neural network can fit well along with the loading paths but often fails to obey the laws of physics, due to a lack of experimental data and understanding of the hidden physical  mechanism. In this paper, we propose a thermodynamic consistent neural network (TCNN) approach to build a data-driven model of the TSR with sparse experimental data. The TCNN leverages recent advances in physics-informed neural networks (PINN) that encode prior physical information into the loss function and efficiently train the neural networks using automatic differentiation. We investigate three thermodynamic consistent principles, i.e., positive energy dissipation, steepest energy dissipation gradient, and energy conservative loading path. All of them are mathematically formulated and embedded into a neural network model with a novel defined loss function. A real-world experiment demonstrates the superior performance of TCNN, and we find that TCNN provides an accurate prediction of the whole TSR surface and significantly reduces the violated prediction against the laws of physics. 

\end{abstract}

\section{Introduction}
Traction-separation relations play a key role in understanding the mechanical behavior of a material interface undergoing openings and predicting interfacial failures under complex loading conditions \cite{park2011cohesive,wu2016determining,yang2019rate,wu2019simultaneous}. However, the entire process is tedious, complex, and unreliable due to three main problems: (1) there has not been a universal and robust approach that can extract the TSR from the cohesive zone using far-field measurements; (2) there has not been a unified approach to model TSR directly from the experimental data; and (3) the experimental design is typically pre-designed, which cannot provide efficient coverage for the testing space \cite{park2011cohesive}. Therefore, it is urgent to develop a data-driven approach to model interfacial TSR, which allows us to effectively learn from sparse experimental data and comply with thermodynamic consistency \cite{wang2019meta}. 

Deep learning has achieved remarkable success in diverse applications \cite{goodfellow2016deep,lecun2015deep} including computer vision and natural language processing, but its use in real-world engineering fields with small data is limited. For the TSR problem, a neural network can fit well along the loading paths but often fails to obey physical laws, due to a lack of experimental data and understanding of the inherent mechanism. To this end, we seek a physics-informed approach that enables us to encode physical laws as prior information into deep learning models, which can mitigate the issue caused by a lack of data \cite{zhang2018quantification,zhang2018effect}. Recent advances in physics-informed neural networks (PINN) \cite{raissi2019physics,raissi2017physics,raissi2017physicsII} that have been used in a wide range of engineering applications including fluid mechanics \cite{mao2020physics, raissi2020hidden}, bio-medical engineering \cite{kissas2020machine}, nanophotonics \cite{chen2020physics,zhang2020directional} and computational materials science\cite{zhang2020robust,liu2020monte}, may bring an opportunity to address this challenge. PINN aims at solving supervised learning tasks while respecting any given law of physics described by general nonlinear PDE. The trained neural networks represent a class of data-efficient approximators that naturally encode underlying physical laws as prior information. This important feature of PINN enables solving inverse problems with limited data observations \cite{zhang2019quantifying}. However, it is a non-trivial task to simply use the PINN for the TSR problem because there are three challenges: (1) the law of physics hidden in TSR is complicated and can not be explicitly described by PDE governing equations; (2) the thermodynamic consistency in TSR is more abstract and difficult to be extracted than typical PDE-based governing equations, such as boundary conditions in PINN; and (3) the prior information from thermodynamic consistency needs to be formulated and implemented into neural network models in a rigorous and data-driven approach. 
% \begin{itemize}  [leftmargin=10pt]
%     \item The law of physics hidden in TSR is complicated and can not be explicitly described by partial differential equation (PDE) governing equations;
%     \item The thermodynamic consistency in TSR is more abstract and difficult to be extracted than typical PDE-based governing equations, such as boundary conditions in PINN;
%     \item The prior information from thermodynamic consistency needs to be formulated and implemented into neural network models in a rigorous and data-driven approach.
% \end{itemize}
% (1) the law of physics hidden in TSR is complicated and can not be explicitly described by PDE governing equations; (2) the thermodynamic consistency in TSR is more abstract and difficult to be extracted than typical PDE-based governing equations, such as boundary conditions in PINN; and (3) the prior information from thermodynamic consistency needs to be formulated and implemented into neural network models in a rigorous and data-driven approach. 
To address these challenges, our core contributions in this paper can be summarized as follows: 

\begin{itemize}  [leftmargin=10pt]
    \item We propose a novel thermodynamic consistent neural network (TCNN) approach to model the material interface mechanics with sparse experimental data 
    \item We extract three thermodynamic consistency principles, i.e., positive energy dissipation, steepest energy dissipation gradient, and energy conservative loading path, from complex TSR problems
    \item We formulate the physical knowledge mathematically, encode the prior information as physics constraints that are then embedded into a neural network model with a new loss function
\end{itemize}

\section{Data-driven modeling of TSR using TCNN}
\begin{figure}[h!]
     \centering
     \includegraphics[width=0.98\textwidth]{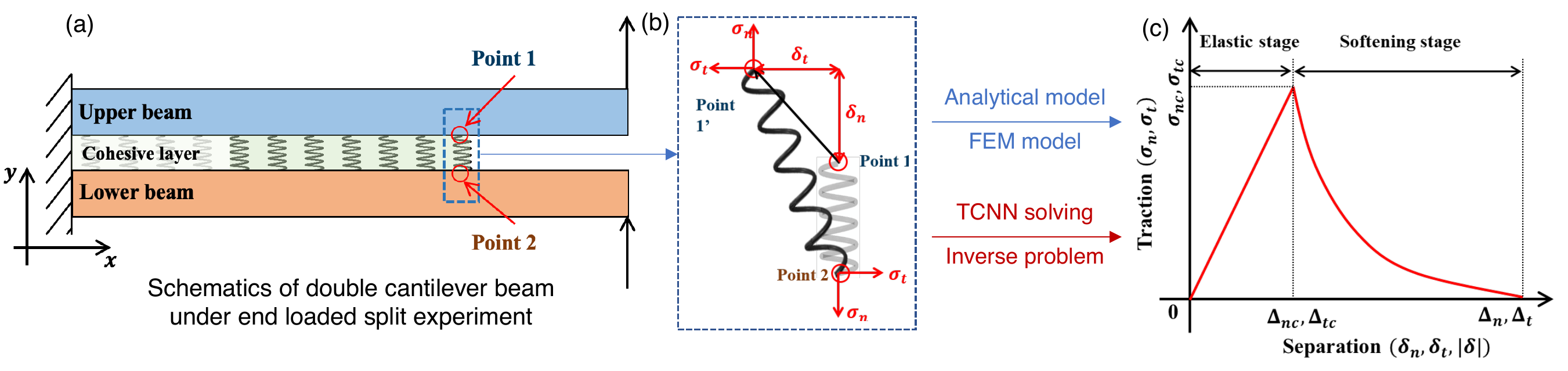}
    \caption{Illustration of TSR problem: (a) Schematics of a double cantilever beam under end loaded split experiment, (b) traction and separations for one spring and (c) a typical TSR curve. }
    \label{fig:TSR}

\end{figure}

{\bf Interfacial traction separation relations (TSR)} \quad For a 2-layer structure undergoing interfacial fracture process as shown in Figure \ref{fig:TSR} (a), a cohesive layer in between these two evolved substrates provides tractions. This cohesive layer is assumed to be homogeneous, which enables us to consider it as an assembly of identical “springs” connecting the two layers. The interfacial fracture process is then reproduced with the elongation and failure process of each “spring” along the interface. For a stretched spring connecting the upper and lower layer, the normal and tangential tractions ($\sigma_n, \sigma_t)$ change correspondingly with the normal and tangential separations ($\delta_n, \delta_t$), which are defined by the change of the relative distances between the end points of the “spring”. Depending on the loading and boundary conditions, the ratio between the tangential and normal separations (i.e., the mode-mix) varies. To quantitatively describe this relation, the vectorial separation is defined as the Euclidean norm of normal and tangential separation components, shown as $|\delta|$ in Figure \ref{fig:TSR}. The mode mix is represented by the phase angle, which is defined as the arctangent of the ratio between normal and tangential separations, shown as $\phi$ in Figure \ref{fig:TSR}. Similarly, the vectorial traction is defined as $|\sigma|$ in Figure \ref{fig:TSR}. With these definitions, the relation between tractions ($\sigma_n, \sigma_t$) and separations ($\delta_n, \delta_t$) is defined as traction separation relation (TSR) which constitutes the interfacial mechanical property. 

\begin{figure}[h!]
     \centering
     \includegraphics[width=0.98\textwidth]{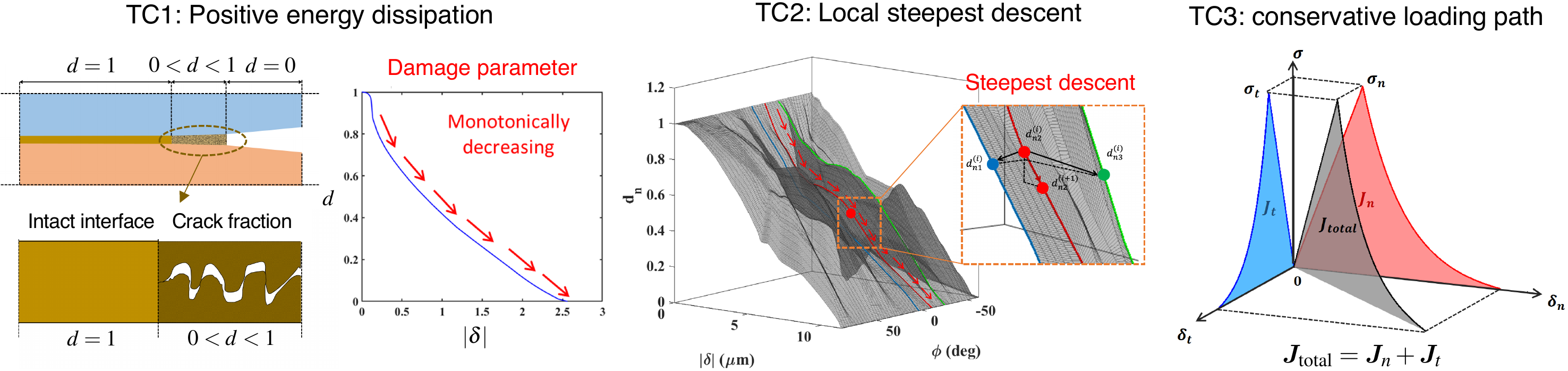}
    \caption{Three thermodynamic consistent principles for modeling TSR}
    \label{fig:TCs}
\end{figure}

{\bf Thermodynamic consistent principles} \quad We extract three principles and discuss them as follows:

\begin{itemize}  [leftmargin=10pt]
    % \item $\textup{TC1}$: {\em Positive energy dissipation}. Damage mechanics at the interface is the foundation for cohesive modeling and the cohesive zone can be represented as the partial fracture at the interface. The total interfacial fracture resistant energy can then be described as a damaging manner $\Gamma = (1-d_n)\Gamma_n + (1-d_t) \Gamam_t$, where $\Gamma$ is the total interfacial toughness, $\Gamma_n$ and $\Gamma_t$ are the normal and shear toughness for the intact interface, $d_n$ and $d_t$ are the damage parameters ranging from $[0,1]$, and can be defined by $d_n(\delta_n) = 1 - J_n(\delta_n)/{\Gamma_n}$ and $d_t(\delta_t) = 1 - {J_t(\delta_t)}/{\Gamma_t}$ based on the $J$-integrals $J_n(\delta_n) = \int_{0}^{x=\delta_n} \sigma_n(x)dx$ and $J_t(\delta_t) = \int_{0}^{x=\delta_t}  \sigma_t(x)dx$ and toughness $\Gamma_n = \max(J_n)$ and $\Gamma_t = \max(J_t)$. For a monotonic loading during the experiment, the energy dissipation for the interfacial delamination should be positive $D = \Gamma_n\dot{d}_n + \Gamma_t\dot{d}_t \ge 0$. This implies that the rate of the damage parameters should be positive for the monotonic loading, $\partial{d_n}/\partial{\delta_n} \ge 0$ and $\partial{d_t}/\partial{\delta_t} \ge 0$. 
    % \begin{equation}
    %     d_n(\delta_n) = 1 - \frac{J_n(\delta_n)}{\Gamma_n}, \quad J_n(\delta_n) = \int_{0}^{x=\delta_n} \sigma_n(x)dx
    % \end{equation}
    %     \begin{equation}
    %     d_t(\delta_t) = 1 - \frac{J_t(\delta_t)}{\Gamma_t}, \quad J_t(\delta_t) = \int_{0}^{x=\delta_t} \sigma_t(x)dx
    % \end{equation}
    
    \item $\textup{TC1}$: {\em Positive energy dissipation}. Damage mechanics at the interface is the foundation for cohesive modeling and the cohesive zone can be represented as the partial fracture at the interface. The total interfacial fracture resistant energy can then be described as a damaging manner 
    \begin{equation}
        \Gamma = (1-d_n)\Gamma_n + (1-d_t) \Gamma_t
    \end{equation}
     where $\Gamma$ is the total interfacial toughness, $\Gamma_n$ and $\Gamma_t$ are the normal and shear toughness for the intact interface, $d_n$ and $d_t$ are the damage parameters ranging from $[0,1]$, and can be defined by 
    \begin{equation}
        d_n(\delta_n) = 1 - {J_n(\delta_n)}/{\Gamma_n}, \quad d_t(\delta_t) = 1 - {J_t(\delta_t)}/{\Gamma_t}
        % d_n(\delta_n) = 1 - \frac{J_n(\delta_n)}{\Gamma_n}, \quad d_t(\delta_t) = 1 - \frac{J_t(\delta_t)}{\Gamma_t}
    \end{equation}
    where $\Gamma_n = \max(J_n)$ and $\Gamma_t = \max(J_t)$ are normal and tangential toughness respectively. $J_n$ and $J_t$ are normal and tangential $J$-integrals which are defined as 
        \begin{equation}
         J_n(\delta_n) = \int_{0}^{\delta_n} \sigma_n(x)dx, \quad J_t(\delta_t) = \int_{0}^{\delta_t} \sigma_t(x)dx \label{eq:j-integral}
    \end{equation}
    For a monotonic loading during the experiment, the energy dissipation for the interfacial delamination should be positive 
    \begin{equation}
    D = \Gamma_n\dot{d}_n + \Gamma_t\dot{d}_t \ge 0. 
    \end{equation}
    This implies that the rate of the damage parameters should be positive for the monotonic loading,
    \begin{equation}
        \frac{\partial{d_n}}{\partial{\delta_n}} \ge 0, \quad \frac{\partial{d_t}}{\partial{\delta_t}} \ge 0
    \end{equation}

    \item $\textup{TC2}$: {\em Steepest energy dissipation gradient}. In addition to the energy dissipation, the dissipation rate should reach the local maximum when the interfacial separation reaches local maximum, 
    \begin{equation}
        \left| \frac{\partial d}{\partial \delta_{\max}}\right|_{\phi = \phi_0} = \max \left|\frac{\partial d}{\partial \bm \delta} \right|
    \end{equation}
    where $\delta_{\max}$ is the vecorial direction towards the largest increment of interfacial separation given a mode-mix phase angle $\phi = \arctan(\delta_t/\delta_n)$. This ensures the fastest energy dissipation follows the fixed loading path. This can be illustrated by damage parameters versus the total separation and the phase angle in Figure \ref{fig:TCs}. The steepest descend will be guaranteed when the projected total separation for each monotonic loading step is the largest, which is given by  
    \begin{equation}
        \left(\frac{\partial d_n}{\partial \delta} \right)_{\phi=\phi_0, \delta = |\delta|} < \left(\frac{\partial d_n}{\partial \delta} \right)_{\phi \neq\phi_0, \delta \neq |\delta|}, \quad   \left(\frac{\partial d_t}{\partial \delta} \right)_{\phi=\phi_0, \delta = |\delta|} < \left(\frac{\partial d_t}{\partial \delta} \right)_{\phi \neq\phi_0, \delta \neq |\delta|}  
    \end{equation}
    
    \item $\textup{TC3}$: {\em Energy conservative loading path}. The last physics constrain is the fulfillment of energy conservation law, which is the energy dissipation along the vectorial path should equal to the sum of energy dissipation in both normal and tangential direction, as shown in Figure \ref{fig:TCs}. This requires the ratio between the normal tangential stress should be equal to those of separations, i.e., 
    \begin{equation}
        \bm J_{\textup{total}} (\sigma_n, \sigma_t, \delta_n, \delta_t) = \bm J_n(\sigma_n, \delta_n) + \bm J_{t} (\sigma_t, \delta_t)
     \end{equation}
     Thus we have 
     \begin{equation}
     \frac{d \sigma_t} {d \sigma_n} = \frac{d \delta_t}{d \delta_n} = \tan \phi
     \end{equation}
     
    % \item $\textup{TC3}$: {\em Energy conservative loading path}. The last physics constrain is the fulfillment of energy conservation law, which is the energy dissipation along the vectorial path should equal to the sum of energy dissipation in both normal and tangential direction, $\bm J_{\textup{total}} (\sigma_n, \sigma_t, \delta_n, \delta_t) = \bm J_n(\sigma_n, \delta_n) + \bm J_{t} (\sigma_t, \delta_t)$, as shown in Figure \ref{fig:TCs}. This requires the ratio between the normal tangential stress should be equal to those of separations, i.e., 
    % \begin{equation}
    %     \frac{d \sigma_t} {d \sigma_n} = \frac{d \delta_t}{d \delta_n} = \tan \phi
    %  \end{equation}

\end{itemize}
In review of all three thermodynamics consistent conditions, TC1 and TC2 are based on the irreversible nature of the energy dissipation as stated in the second law of thermodynamics, thus are the stronger constrains. TC3 condition is a relatively weaker condition than TC1 and TC2, because it is based on the assumption that the delamination occurs at the interface with no frictional energy dissipation, which may not be true based on some of the experiment observations \cite{wu2019simultaneous}.

\begin{figure}[h!]
% \vspace{-0.2cm}
     \centering
     \includegraphics[width=0.98\textwidth]{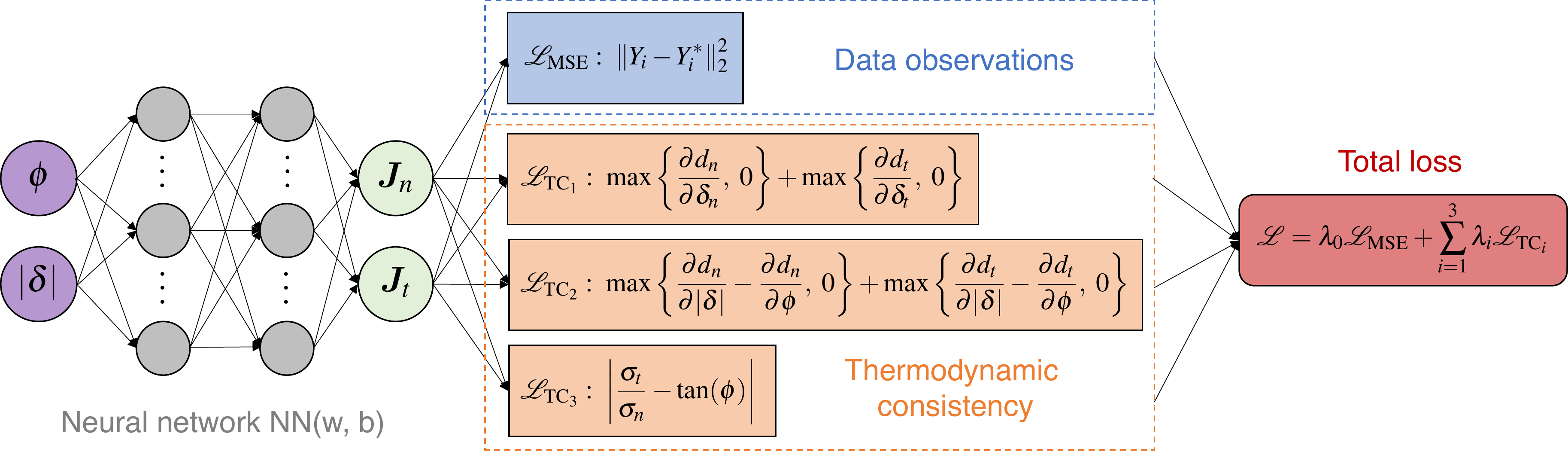}
    \caption{Schematic of a thermodynamic consistent neural network (TCNN) for solving TSR problems. The total loss consists of the loss from data observations and thermodynamic consistency.}
    \label{fig:TCNN}
\end{figure}

{\bf Thermodynamic consistent neural network (TCNN)} \quad We propose a TCNN framework by embedding the thermodynamic consistency principles as physical constraints into deep neural networks. Figure \ref{fig:TCNN} presents the schematic of a TCNN for solving TSR prediction with three thermodynamic consistent constraints. The inputs of the DNN model are separation norm $|\delta|$ and phase angle $\phi$ as defined in Figure \ref{fig:TSR}, and the outputs are normal and tangential $J$-integrals ($J_n, J_t$), as defined in Eq.~\eqref{eq:j-integral}. The goal is to predict the $J$-integral surfaces using trained DNN model with collected experimental data. 
\begin{itemize} [leftmargin=10pt]
    \item $\mathcal{L}_{\textup{MSE}}$: a regular expected MSE loss to measure the mismatch with the given data observations $Y^*=(J_n^*, J_t^*)$, which is defined as 
    \begin{equation}
        \mathcal{L}_{\textup{MSE}} = \frac{1}{z} \sum_{i=1}^z \left(\left \| Y_i - Y_i^{*}\right\|_2^2 \right)
    \end{equation}
    where $i=1,..,z$ is the number of data. 
    
    \item $\mathcal{L}_{\textup{TC}_1}$: TC1 is imposed on constraint paths separately on normal and tangential output $J$-integral surfaces. We use \texttt{max} function to compare the values so that only the positive part is added to the loss,
    \begin{equation}
       \mathcal{L}_{\textup{TC}_1} = \frac{1}{2m} \sum_{j=1}^m \left( \max_{i \in [1, z-1]}\left\{\frac{d_n^{(i+1,j)} - d_n^{(i,j)}}{\delta_n^{(i+1,j)} - \delta_n^{(i,j)}}, \ 0\right\} + \max_{i \in [1, z-1]}\left\{\frac{d_t^{(i+1,j)} - d_t^{(i,j)}}{\delta_t^{(i+1,j)} - \delta_t^{(i,j)}}, \ 0\right\} \right)
    \end{equation}
    
    \item $\mathcal{L}_{\textup{TC}_2}$: TC2 constraints the gradient vector direction along the fixed phase angle loading paths
    \begin{equation}
        \mathcal{L}_{\textup{TC}_2} = \frac{1}{2mz} \left(\sum_{i=1}^{z-1} \sum_{j=1}^{m-1} \max \left\{\frac{\partial d_n^{(i,j)}}{\partial |\delta|} - \frac{\partial d_n^{(i,j)}}{\partial \phi}  , \ 0 \right\} + \max \left\{\frac{\partial d_t^{(i,j)}}{\partial |\delta|} - \frac{\partial d_t^{(i,j)}}{\partial \phi}  , \ 0 \right\} \right)
    \end{equation}
    
    \item $\mathcal{L}_{\textup{TC}_3}$: Different from TC1 and TC2 constraints, TC3 condition is an equality constraint,  
    \begin{equation}
        \mathcal{L}_{\textup{TC}_3} = \frac{1}{mz}\sum_{j=1}^m \sum_{i=1}^z \left| \frac{\sigma_t^{(i,j)}}{\sigma_n^{(i,j)}} -\tan(\phi^{(j)}) \right|
    \end{equation}
    
    \item $\mathcal{L}$: the total loss is defined by the weighted summation of each loss functions   
    \begin{equation}
        \mathcal{L} = \lambda_0 \mathcal{L}_{\textup{MSE}} + \lambda_1 \mathcal{L}_{\textup{TC}_1}  + \lambda_2 \mathcal{L}_{\textup{TC}_2} + \lambda_3 \mathcal{L}_{\textup{TC}_3}  
    \end{equation}
    where $\lambda_i, i=0,1,2,3$ refer to the weights of the loss functions and satisfies $\sum_{i=0}^{3} \lambda_i=1$.
\end{itemize}

\section{Experiments and results}
\begin{figure}[h!]
     \centering
     \includegraphics[width=0.6\textwidth]{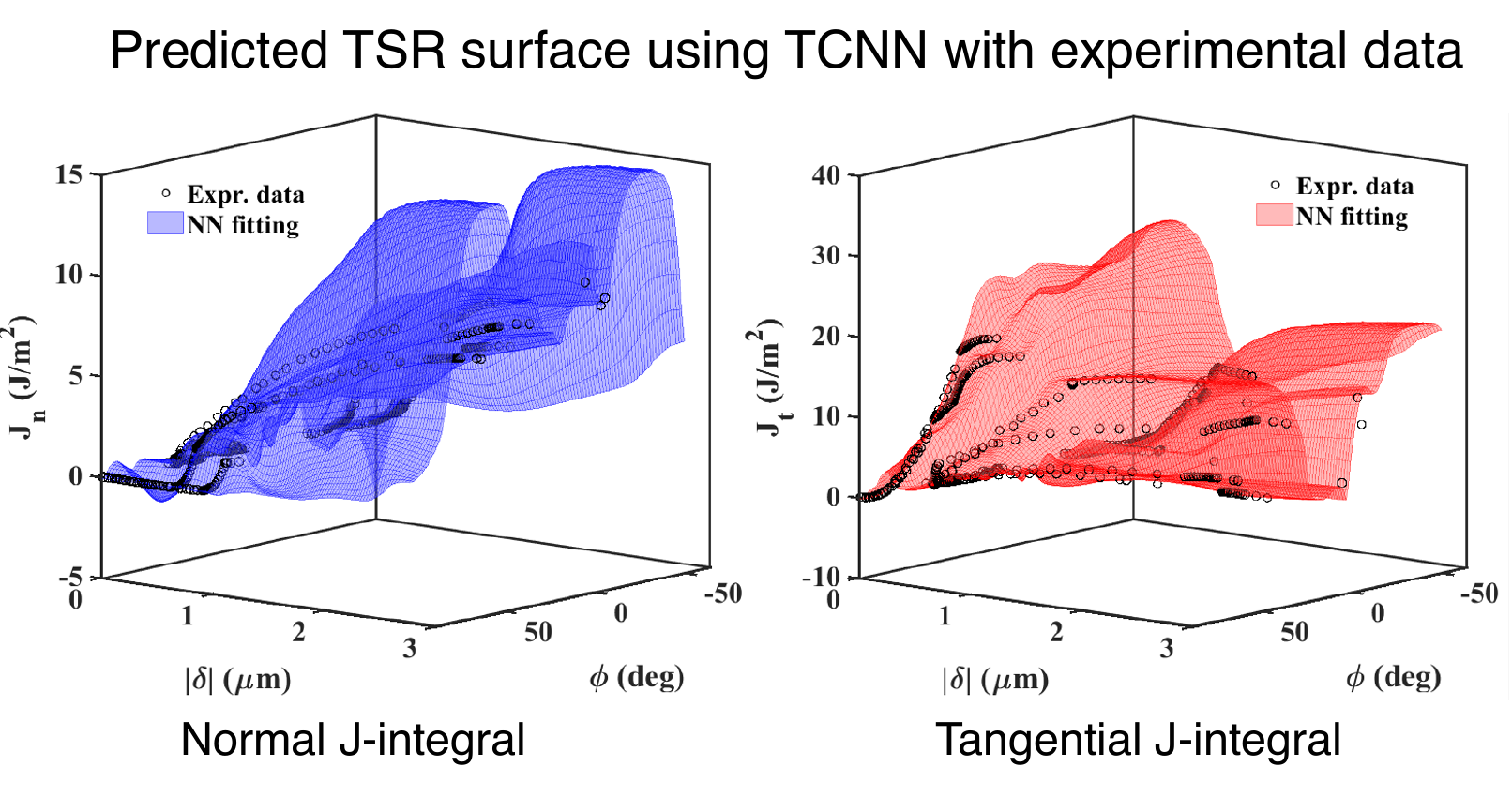}
    \caption{TCNN performance on TSR data: predicted TSR surface using experimental data}
    \label{fig:results1}
\end{figure}

\begin{figure}[h!]
\vspace{-0.2cm}
     \centering
     \includegraphics[width=0.6\textwidth]{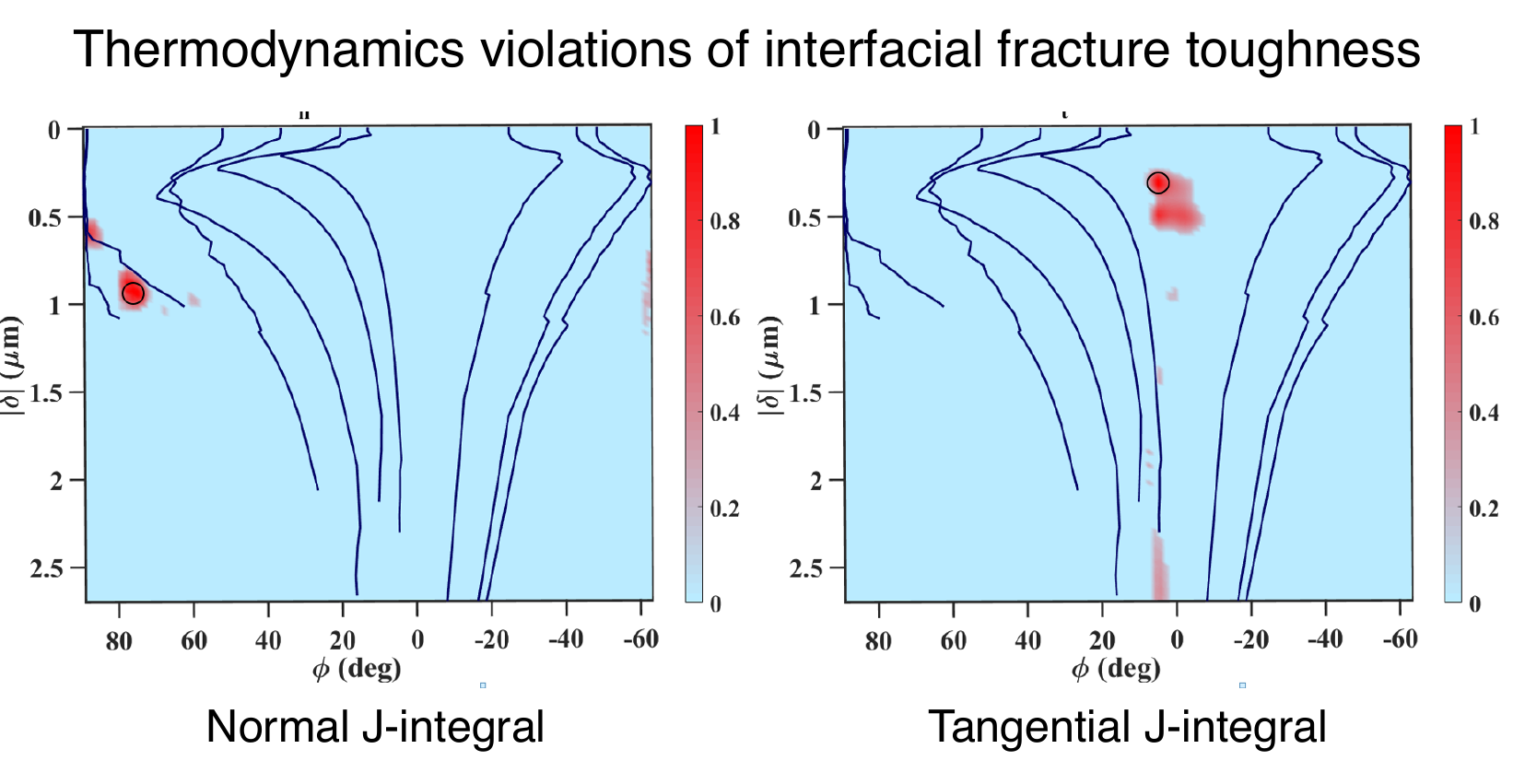}
    \caption{Thermodynamics violations of predicted interfacial fracture toughness. Note that the red color represents the thermodynamic violations, blue color means no violations.}
    \label{fig:results2}
\end{figure}

We collected data from real-world experimental measurements where the normal and shear TSR for a silicon/epoxy interface were determined over normal mode-mixes range from -53° to 87.5° using non-symmetric end-loaded split (ELS) and end-notched flexure (ENF) specimens \cite{wu2019simultaneous}. A total 10 loading paths with 236 TSR data points are feed to the TCNN model in which a neural network model is constructed with two layers and there are 60 neurons for each layer, and we use \texttt{tanh} as the activation function and Adam as the optimizer. The hyperparameters are tuned by trust-region Bayesian optimization \cite{eriksson2019scalable}. Figure \ref{fig:results1} and Figure \ref{fig:results2} shows the performance of TCNN on TSR surface using sparse experimental data. Note that the learned TSR model can capture the complex surface even only limited data is given and the thermodynamic violations of interfacial fracture toughness is around 5\% that is much lower than the cases without encoding thermodynamic constraints (typically 20\% - 40\%). 


\section{Conclusion}
In this paper, we propose a thermodynamic consistent neural network method for solving the traction separation relation problems in material interactial mechanics. Three thermodynamic consistency principles are formulated as physics constraints and mathematically imposed to a deep neural network model with a novel loss function. By introducing these prior physical knowledge into the deep learning model, we demonstrate that the TCNN enables to well predict the entire traction surface, and significantly reduce the violated prediction against the laws of thermodynamics.  

\clearpage
\bibliographystyle{plain}
\bibliography{reference}

\begin{thebibliography}{10}

\bibitem{chen2020physics}
Yuyao Chen, Lu~Lu, George~Em Karniadakis, and Luca Dal~Negro.
\newblock Physics-informed neural networks for inverse problems in nano-optics
  and metamaterials.
\newblock {\em Optics Express}, 28(8):11618--11633, 2020.

\bibitem{eriksson2019scalable}
David Eriksson, Michael Pearce, Jacob Gardner, Ryan~D Turner, and Matthias
  Poloczek.
\newblock Scalable global optimization via local bayesian optimization.
\newblock In {\em Advances in Neural Information Processing Systems}, pages
  5496--5507, 2019.

\bibitem{goodfellow2016deep}
Ian Goodfellow, Yoshua Bengio, Aaron Courville, and Yoshua Bengio.
\newblock {\em Deep learning}, volume~1.
\newblock MIT press Cambridge, 2016.

\bibitem{kissas2020machine}
Georgios Kissas, Yibo Yang, Eileen Hwuang, Walter~R Witschey, John~A Detre, and
  Paris Perdikaris.
\newblock Machine learning in cardiovascular flows modeling: Predicting
  arterial blood pressure from non-invasive 4d flow mri data using
  physics-informed neural networks.
\newblock {\em Computer Methods in Applied Mechanics and Engineering},
  358:112623, 2020.

\bibitem{lecun2015deep}
Yann LeCun, Yoshua Bengio, and Geoffrey Hinton.
\newblock Deep learning.
\newblock {\em nature}, 521(7553):436--444, 2015.

\bibitem{liu2020monte}
Xianglin Liu, Jiaxin Zhang, Junqi Yin, Sirui Bi, Markus Eisenbach, and Yang
  Wang.
\newblock Monte carlo simulation of order-disorder transition in refractory
  high entropy alloys: a data-driven approach.
\newblock {\em Computational Materials Science}, 187:110135.

\bibitem{mao2020physics}
Zhiping Mao, Ameya~D Jagtap, and George~Em Karniadakis.
\newblock Physics-informed neural networks for high-speed flows.
\newblock {\em Computer Methods in Applied Mechanics and Engineering},
  360:112789, 2020.

\bibitem{park2011cohesive}
Kyoungsoo Park and Glaucio~H Paulino.
\newblock Cohesive zone models: a critical review of traction-separation
  relationships across fracture surfaces.
\newblock {\em Applied Mechanics Reviews}, 64(6), 2011.

\bibitem{raissi2019physics}
Maziar Raissi, Paris Perdikaris, and George~E Karniadakis.
\newblock Physics-informed neural networks: A deep learning framework for
  solving forward and inverse problems involving nonlinear partial differential
  equations.
\newblock {\em Journal of Computational Physics}, 378:686--707, 2019.

\bibitem{raissi2017physics}
Maziar Raissi, Paris Perdikaris, and George~Em Karniadakis.
\newblock Physics informed deep learning (part i): Data-driven solutions of
  nonlinear partial differential equations.
\newblock {\em arXiv preprint arXiv:1711.10561}, 2017.

\bibitem{raissi2017physicsII}
Maziar Raissi, Paris Perdikaris, and George~Em Karniadakis.
\newblock Physics informed deep learning (part ii): Data-driven discovery of
  nonlinear partial differential equations.
\newblock {\em arXiv preprint arXiv:1711.10566}, 2017.

\bibitem{raissi2020hidden}
Maziar Raissi, Alireza Yazdani, and George~Em Karniadakis.
\newblock Hidden fluid mechanics: Learning velocity and pressure fields from
  flow visualizations.
\newblock {\em Science}, 367(6481):1026--1030, 2020.

\bibitem{wang2019meta}
Kun Wang and WaiChing Sun.
\newblock Meta-modeling game for deriving theory-consistent,
  microstructure-based traction--separation laws via deep reinforcement
  learning.
\newblock {\em Computer Methods in Applied Mechanics and Engineering},
  346:216--241, 2019.

\bibitem{wu2016determining}
Chenglin Wu, Shravan Gowrishankar, Rui Huang, and Kenneth~M Liechti.
\newblock On determining mixed-mode traction--separation relations for
  interfaces.
\newblock {\em International Journal of Fracture}, 202(1):1--19, 2016.

\bibitem{wu2019simultaneous}
Chenglin Wu, Rui Huang, and Kenneth~M Liechti.
\newblock Simultaneous extraction of tensile and shear interactions at
  interfaces.
\newblock {\em Journal of the Mechanics and Physics of Solids}, 125:225--254,
  2019.

\bibitem{yang2019rate}
Tianhao Yang, Xingwei Yang, Rui Huang, and Kenneth~M Liechti.
\newblock Rate-dependent traction-separation relations for a silicon/epoxy
  interface informed by experiments and bond rupture kinetics.
\newblock {\em Journal of the Mechanics and Physics of Solids}, 131:1--19,
  2019.

\bibitem{zhang2019quantifying}
Dongkun Zhang, Lu~Lu, Ling Guo, and George~Em Karniadakis.
\newblock Quantifying total uncertainty in physics-informed neural networks for
  solving forward and inverse stochastic problems.
\newblock {\em Journal of Computational Physics}, 397:108850, 2019.

\bibitem{zhang2020directional}
Jiaxin Zhang, Sirui Bi, and Guannan Zhang.
\newblock A directional gaussian smoothing optimization method for
  computational inverse design in nanophotonics.
\newblock {\em Materials \& Design}, 197:109213.

\bibitem{zhang2020robust}
Jiaxin Zhang, Xianglin Liu, Sirui Bi, Junqi Yin, Guannan Zhang, and Markus
  Eisenbach.
\newblock Robust data-driven approach for predicting the configurational energy
  of high entropy alloys.
\newblock {\em Materials \& Design}, 185:108247, 2020.

\bibitem{zhang2018effect}
Jiaxin Zhang and Michael~D Shields.
\newblock The effect of prior probabilities on quantification and propagation
  of imprecise probabilities resulting from small datasets.
\newblock {\em Computer Methods in Applied Mechanics and Engineering},
  334:483--506, 2018.

\bibitem{zhang2018quantification}
Jiaxin Zhang and Michael~D Shields.
\newblock On the quantification and efficient propagation of imprecise
  probabilities resulting from small datasets.
\newblock {\em Mechanical Systems and Signal Processing}, 98:465--483, 2018.

\end{thebibliography}

\end{document}